\documentclass{rjparticle}
\newcommand{\be}{\begin{equation}}
\newcommand{\ee}{\end{equation}}
\begin{document}
\title{Slow roll inflation: a somehow different perspective}
\author{Cristiano Germani}
\affil{Arnold Sommerfeld Center, Ludwig-Maximilians-University,\\Theresienstr. 37, 80333 Munich, Germany}
\affil{Email: cristiano.germani@lmu.de}

\maketitle

\begin{abstract}
In this note we point out that, contrary to the standard point of view, slow roll inflation is due to high gravitational friction. We show that the requirement of slow roll coincides with the requirement of a flat scalar field potential in the case of minimally coupled scalar field.
In this sense, the search for a successful inflationary theory may be more fruitful by shifting the focus on models with high gravitational friction. We review then a gravitational mechanism, the so called ''Gravitationally Enhanced Friction" mechanism, such that high gravitational friction is dynamically generated during inflation allowing even steep (i.e. non-flat) scalar potential to inflate.
\end{abstract}
\section{Inflation}
At large scales, the Universe looks extremely homogeneous isotropic and flat. This can be dynamically obtained
by invoking an inflationary period in the early stages of the Universe evolution. The idea is very simple, it basically plays with the fact that any surface is locally flat homogeneous and isotropic:
 
Suppose the Universe, in the very past, was not homogeneous, not isotropic and spatially curved. Nevertheless, in any small patch, it would look like homogeneous isotropic and flat. Inflation, is literarily a fast (exponential) blowing of this small patch so to obtain, out of it, the today observable ``large'' Universe. Precisely, inflation is realized whenever there is an exponential run away of each point ($i,j$) of the small patch, i.e.
\be\label{universe}
{\vec R}_{ij}(t)=e^{Ht}\vec{R}_{ij}(t_0)\ ,
\ee
where ${\vec R}_{ij}(t)$ is the relative physical distance\footnote{Note that in General Relativity the physical distance differs from the coordinate distance.}, $H$ is a constant, $t$ is the cosmic time and $t_0$ is some initial time.
To get an intuition of why inflation is produced by gravitational friction, we will now extensively use a Newtonian-like description of the Universe expansion. 

Let us fix the coordinates to be centered in one of the points. If only gravity is the force responsible for the Universe evolution we would have, from Newton's law\footnote{We suppress from now on the $i,j$ indices and we consider units where the reduced Planck scale is $1$.}
\be
\ddot R=-\frac{M(R)}{R^2}\ ,
\ee
where $R=|\vec{R}|$ and $M(R)$ is the mass (self gravitating energy) of the Universe. By using (\ref{universe}) we easily see that
\be
H^2=-\frac{M(R)}{R^3}\ ,
\ee
or
\be\label{4}
M(R)=-H^2 R^3\sim - H^2 V_3\ ,
\ee
where $V_3$ is the three-dimensional volume. Eq. (\ref{4}) implies that an inflating Universe is sourced by a constant ``negative'' gravitating energy per unit volume. Although this looks odd in the context of Newtonian theory, it is not in the context of General Relativity (GR). Let us consider a perfect fluid of energy density $\rho$ and an isotropic pressure $p$. In GR, any form of energy gravitates, in particular $p$ is a form of energy. Thus, if we discard negative energy density configurations, a negative gravitating energy can be obtained with a negative pressure, as we shall see. 

The gravitational potential $U$ is an energy and therefore it is related to time translation of the system. The pressure instead modifies space translations. Therefore, a metric describing a homogeneous and isotropic universe driven by a perfect fluid may be written in ``Newtonian'' coordinates (see \cite{nicolis}), as
\be\label{newt}
ds^2=-(1-2U)dt^2+(1-2\Psi)dx\cdot dx\ .
\ee
From the Einstein equation $G_{\alpha\beta}=T_{\alpha\beta}=(\rho+p)u_\alpha u_\beta+p g_{\alpha\beta}$, where $u_{\alpha}=\delta_\alpha^t$, we have the following (independent) $tt$ and $ii$ components
\begin{eqnarray}
\nabla^2\Psi=\frac{\rho}{2}\ ,\cr
6\ddot\Psi+2\nabla^2(U-\Psi)=3p\ .
\end{eqnarray}
Combining the previous equations and considering a constant energy density, we finally obtain the General Relativistic Poisson equation
\be
\nabla^2 U=\frac{1}{2}\left(\rho+3p\right)\ .
\ee
For $\rho>0$ and $\rho+3p<0$ we can get inflaton if $p$ is constant and negative, as anticipated. In the Newtonian limit, in which the pressure is much smaller than the energy density ($\rho\gg p$), we obtain the Poisson equation for Newtonian gravity.

\section{Slow Roll is High Friction}

An important constraint for inflation is that it should end in order to form Galaxies. Therefore, it cannot be generated by a bare constant gravitational energy density. A natural Inflaton candidate is thus a scalar field that can interpolate from an inflating phase ($\rho+3p<0$) to a deflating phase ($\rho+3p>0$). A byproduct of this is that the Universe cannot undergo to an exact exponential expansion.

The first issue we want to address is to get a roughly constant gravitational energy out of the scalar field. Suppose to have a scalar field with positive definite potential $V$. For simplicity and to get connection to the Newtonian intuition we just consider a mass term $V=\frac{1}{2}m^2\phi^2$, however, our discussion can be easily generalized. The energy density and pressure can only be obtained by a combination of kinetic and potential energies. A sufficient condition to have almost constant gravitational energy per unit volume, is to have (separately) almost constant kinetic and potential energies per unit volume of the scalar field. 

Let us start from the Newtonian-like evolution of a massive scalar field $\phi$:
\be\label{original}
\ddot\phi=-m^2\phi\ .
\ee
The energy density of this system is given by the first time integral of (\ref{original}), i.e.
\be
\rho=\frac{1}{2}\dot\phi^2+\frac{1}{2}m^2\phi^2\ .
\ee
It is clear that to have an almost constant kinetic and potential energies, the kinetic term should be very small compared to the potential term. From the equation (\ref{original}), this would imply $m^2\phi\simeq 0$. This last condition, does not allow inflation as $U\sim 0$. Generically though, one may have $V'\sim 0$ but $V>0$. However, here, there is no end of Inflation. We therefore conclude that, in order to have a successful inflation, (\ref{original}) must be modified. We can now use again our intuition from classical mechanics. 

Let us consider a ball falling down in a gravitational potential. The gravitational potential would make the ball accelerate and therefore would break quickly the condition of constancy of kinetic and potential energies. We can now change the experiment. Let the ball falling down in a viscous medium. In the high friction limit, the ball would fall very slowly so to drastically reduce the kinetic energy. Mathematically, one may then drastically reduce the kinetic energy of the scalar field by introducing a friction term, in the evolution equation of the scalar field, that dominates over the acceleration. A friction term is proportional to the velocity, so the scalar field evolution we look for is
\be
D\dot\phi\simeq -m^2\phi\ ,
\ee
where $D$ represents the strength of the (at the moment constant) friction. For an initial $\phi_0\neq 0$ the solution of this equation is
\be\label{sf}
\phi\simeq \phi_0 e^{-\frac{m^2}{D}t}\ .
\ee
The contribution of the friction to the gravitational energy density would just come from the first integral of the friction term, i.e. $D\left(\phi-\phi_0\right)$, which is very small for $\phi\simeq \phi_0$.
 
Comparing the scalar field variation (\ref{sf}) with the Universe expansion (\ref{universe}), we find that for
\be\label{condition}
\frac{m^2}{D}\ll H\ ,
\ee
the scalar field is roughly constant during the Universe evolution, and in turn implies slow roll, i.e. 
\be
\dot\phi\ll V\ .
\ee
What we then learned is that no matter the value of the mass, for large enough friction we get a slow rolling scalar field. This is the main point of this note. Slow roll is indeed strictly speaking dominated by friction and not by flatness of the potential. However, as we shall see, in the minimally coupled scalar field case, high friction and flatness of the potential is a synonym.

So far so good, however, we still need to consider a non-constant friction term, in order to end Inflation. In particular we want a friction term to be efficient only at high (gravitational) energies. As we said Inflation cannot be an exact exponential expansion, this means that $H=H(t)$ with the requirement of a slow variation during inflation, i.e. $-\dot H/H^2\equiv\epsilon\ll 1$. Inflation will then end when $\epsilon\sim 1$. We can then promote the friction to be slowly varying by taking it as a function of $H$, in other words we require $D=D(H)$ and in particular $\partial_H D>0$ in order to obtain an inefficient friction at low energies.

\section{The Making of Slow Roll}

By dimensional reasons, the simplest friction is $D=\alpha H$ where $\alpha$ is a dimensionless constant of order one. In this case, the condition (\ref{condition}) boils down to ${\cal O}(\phi^2)\gg 1$, or 
\be
{\cal O}\left(\frac{{V'}^2}{V^2}\right)\ll 1\ ,
\ee
where $V'\equiv\partial_\phi V$ and, because of slow roll ($\dot\phi\ll V$), we have used the fact that the gravitational energy density ($H^2$) is of order $V$. Therefore, in the simplest case $D\propto H$, there is a one-to-one correspondence between high friction and flat potential.

Let us then consider the next to simplest model, i.e. a friction depending on some positive power of $H$, $D=\frac{H^n}{M^{n-1}}$, where $n>1$ and $M$ is a mass scale. We will call this class of model the GEF (gravitationally enhanced friction) models. For $n=3$ they are covariantly realized in \cite{cov1,cov}. In this case the condition (\ref{condition}) boils down to
\be
\left(\frac{m}{M}\right)^{n-1}\gg {\cal O}\left[\left(\frac{V'}{V}\right)^{n+1}\right]\ .
\ee
This last inequality teaches us that no matter the curvature of the potential $V$, as long as the mass $M$ is much below the Inflaton mass during inflation, slow roll is obtained.

\section{Covariant realization of gravitational friction}

From now on we abandon the Newtonian description and we use a full GR description. Inflation, is described by a Friedman-Robertson-Walker (FRW) line element\footnote{Note that this line element is related to (\ref{newt}) by a coordinate transformation \cite{nicolis}.}
\be
ds^2=-dt^2+a(t)^2dx\cdot dx\ .
\ee
One can easily see that an inflating background $H={\rm cont}$ as discussed before, means $\frac{\dot a}{a}=H$. This expansion of the spatial metric introduce a friction term in the minimally coupled scalar field thanks to the volume element $\sqrt{-g}=a^3$. In fact, as it is well known, the Klein-Gordon equation is 
\be
\frac{1}{\sqrt{-g}}\partial_\alpha\left(\sqrt{-g}\partial^\alpha\phi\right)=-V'\ .
\ee
Expanding the left hand side we then see that one gets automatically a friction term $D=3H$. The Universe, while expanding, acts as a friction term for the scalar field!

The question is now: how can we get an enhanced gravitational friction without introducing new fields (or degree of freedom)?

As we shall see there is only one possibility \footnote{We are not considering here higher powers of scalar field derivatives. In that case in fact, whenever Inflation is possible, the explanation in terms of friction gets difficult.} namely $D\sim \frac{H^3}{M^2}$. During inflation the Einstein tensor is roughly constant and in particular $G^{tt}\simeq 3H^2$. We can then modify the Klein-Gordon equation as follow \cite{cov1, cov}
\be\label{kg}
g^{\alpha\beta}\nabla_\alpha\nabla_\beta\phi\rightarrow\left(g^{\alpha\beta}-\frac{G^{\alpha\beta}}{M^2}\right)\nabla_\alpha\nabla_\beta\phi\ ,
\ee
where $\nabla_\alpha$ is the covariant derivative. If then, during inflation, $H\gg M$ (high friction regime) we readily obtain the enhanced friction
\be
D\simeq 9\frac{H^3}{M^2}\ .
\ee

The shift in the Klein-Gordon differential operator can be obtained via the (GEF) kinetic Lagrangian
\be\label{action}
A_k=-\frac{1}{2}\int d^4x\sqrt{-g}\left(g^{\alpha\beta}-\frac{G^{\alpha\beta}}{M^2}\right)\partial_\alpha\phi\partial_\beta\phi\ .
\ee
Strictly speaking, during slow roll, the lagrangian (\ref{action}) really describes a time rescaling of the scalar field physical clock 
\be
t\rightarrow \frac{t}{1+3\frac{H^2}{M^2}}\ .
\ee
This re-scaling does the same job of friction by making effectively the time evolution of the scalar field much slower than the time evolution of the inflating Universe. However, in slow roll, whenever $\ddot\phi\ll H\dot\phi$, time rescaling and physical friction are equivalent.

It has been shown \cite{cov1,hord} that by varying the action (\ref{action}) with respect both the scalar field and the metric, no new degrees of freedom appear thanks to the Bianchi identities $\nabla_\alpha G^\alpha_\beta\equiv0$. In particular, $A_k$ is the unique kinetic term for the scalar field which does not propagate more degree of freedom rather than a scalar and a graviton \cite{hord}. Therefore, any other power of $H$ in $D$ should necessarily introduce extra degrees of freedom.

It is interesting to note that the action (\ref{action}) may play a fundamental role for dynamical localization of scalars on braneworlds (for other spins one can also write similar lagrangians) \cite{loc}. One might then obtain the theory (\ref{action}) as a dimensionally reduced theory in braneworlds, this open question is left for future work.

\section{Special properties of the GEF lagrangian}

Let us now embed the lagrangian (\ref{action}) in a full gravitational theory (we now re-introduce the Planck scale for clarity). We then consider the Slotheonic \cite{sloth} theory
\be\label{A}
A=\frac{1}{2}\int d^4 x\sqrt{-g}\left[M_p^2 R(g)-2V(\phi)\right]+A_k\ ,
\ee
where $M_p$ is the planck mass.

In the high friction limit
\be
\frac{\frac{G^{\alpha\beta}}{M^2}\partial_\alpha\phi\partial_\beta\phi}{(\partial\phi)^2}\gg 1\ ,
\ee
and in the slow roll trajectory
\be
\frac{\partial_\alpha\phi\partial_\beta\phi}{M^2 M_p^2}\equiv \epsilon_{\alpha\beta}\ll 1\ ,
\ee
the action (\ref{A}) may be recast (at first order in $\epsilon_{\alpha\beta}$) into \cite{sloth}
\be
A_h=\frac{1}{2}\int d^4 x\sqrt{-h}\left[M_p^2 R(h)-2V(\chi)-(\partial\chi)^2\right]\ ,
\ee
where the Finsler metric $h_{\alpha\beta}$ is defined as
\be\label{fieldred}
h_{\alpha\beta}\equiv g_{\alpha\beta}-\frac{\partial_\alpha\phi\partial_\beta\phi}{M^2 M_p^2}\ ,
\ee
and the canonically normalized field
\be
\chi\equiv\int d\phi\frac{V^{1/2}}{M M_p}\ .
\ee
As an example let us again use quadratic potential for $\phi$ so to obtain in the Finsler theory an effective linear potential
\be
V(\chi)=mMM_p\chi\ .
\ee
We see then that, in this ``Einstein'' frame for the Slotheonic theory in high friction and slow roll limits, the effective field $\chi$ inflates the Universe if and only if $\chi^2\gg M_p^2$, as in the canonical case of $D=3H$. The avoidance of trans-Planckian curvatures then is just recasted into $m\ll M_p$ (in the GEF case in which $M\ll m$\footnote{Actually this condition might be much weaker, dependently on the curvature of the potential $V(\phi)$.}). Linear perturbations around the inflating background may then be easily calculated in this frame with standard techniques \cite{malda}. One can then go back to the original (non-Einstein) frame by the metric field redefinition (\ref{fieldred}) and infers the physically observed spectral index.

\section{Conclusions}

In this note, we showed that Inflation can only be obtained thanks to a high gravitational friction acting on a scalar field rolling down its own potential. We showed that in the minimally coupled case the high friction requirement is equivalent to the requirement of an almost flat potential. We then reviewed a mechanism that increases gravitationally the friction on the scalar field such to make the scalar field slow rolling even on steep potentials. This is achieved by a non-minimal coupling of the kinetic term of the Inflaton to curvatures in a way that does not introduce any higher derivatives and/or new degrees of freedom rather than a scalar and a spin-2 (the graviton).

\begin{acknowledgement}
I wish to thank Yuki Watanabe for important comments on the first draft of this note. I wish to thank the organizers of the BW2011 Workshop ``Particle Physics from TeV to Plank Scale'' for a stimulating conference. I am supported by the  by Alexander Von Humboldt Foundation.

\end{acknowledgement}

\end{document}